# Stabilising a nulling interferometer using optical path difference dithering

Pavel Gabor[1], Bruno Chazelas[1], Frank Brachet[2], Marc Ollivier[1], Michel Decaudin[1], Sophie Jacquinod[1], Alain Labèque[1], and Alain Léger[1]

[1] Institut d'Astrophysique Spatiale, Orsay, France
[2] Centre National d'Etudes Spatiales, Toulouse, France

**Abstract**

*Context.* Nulling interferometry has been suggested as the underlying principle for the *Darwin* and TPF-I exoplanet research missions.
*Aims.* There are constraints both on the mean value of the nulling ratio, and on its stability. Instrument instability noise is most detrimental to the stability of the nulling performance.
*Methods.* We applied a modified version of the classical dithering technique to the optical path difference in the scientific beam.
*Results.* Using only this method, we repeatedly stabilised the dark fringe for several hours. This method alone sufficed to remove the $1/f$ component of the noise in our setup for periods of 10 minutes, typically. These results indicate that performance stability may be maintained throughout the long-duration data acquisitions typical of exoplanet spectroscopy.
*Conclusions.* We suggest that further study of possible stabilisation strategies should be an integral part of *Darwin*/TPF-I research and development.

**Key words.** Nulling interferometry - *Darwin*/TPF-I - optical path stabilisation - optical path difference dithering

## 1. Nulling Interferometry and Exoplanet Research

The method proposed for exoplanet research for the *Darwin* (Karlsson & Kaltenegger 2003, European Space Agency-SCI 12, 2000) and TPF-I (Coulter 2003, JPL Publ. 05-5, 2005) missions is based on nulling interferometry Bracewell (1978) designed to enable IR spectroscopic measurements of exoplanetary atmospheres as well as imagery of extrasolar planetary systems. The challenge is a daunting one with manifold sources of noise Lay (2004): every photon coming from the exoplanet has to be acquired and separated from the noise by all means available. This implies that strategies have to be designed to improve the signal-to-noise ratio (S/N) during the data acquisition stage, while developing efficient algorithms for work with the acquired data. This article comments on the former approach.

Experimental studies of nulling interferometer breadboards (Serabyn 2003; Schmidtlin et al. 2005; Ollivier et al. 2001; Vink et al. 2003; Alcatel 2004; Brachet 2005, etc.) show that even in simple setups, the interference pattern is unstable, drifting with time. Even interferometers breadboarded on an optical bench in the relatively well-controlled laboratory environment (a priori simpler than the actual *Darwin*/TPF-I, with its multiple telescopes rotating in space) display drifts.

Chazelas et al. (2006) suggest "that special attention be given to servo systems... for monitoring key quantities such as the optical path difference (OPD) because these systems [have to] be free of long-term drifts" to obtain the required performance throughout long integration times, e.g. 10 days. Their paper gives a quantitative summary of these effects, using data from Ollivier (1999); Alcatel (2004); Vink et al. (2003), and suggests that servo mechanisms, using the signal itself, be employed to control drifts.

Chazelas et al. (2006) find that the "quality of the null" at a given wavelength and at a given moment in time can be evaluated in terms of the contrast in intensity between two adjacent dark and bright fringes. It can be expressed as the "nulling ratio" (also referred to as "stellar leakage" because it represents the stellar flux that the interferometer tries to cancel) due to the nulling instrument

$$nl(\lambda, t) = \frac{I_{\min}}{I_{\max}}$$

where $I_{\min}$ and $I_{\max}$ stand for the intensity of the onaxis dark fringe and of the offaxis bright fringe, respectively. Chazelas et al. (2006) show that two types of requirements must be met: one requirement is imposed upon the mean value of the nulling ratio $nl(\lambda, t)$, and the other upon its stability.

Taking into account the wavelength dependence of the star/planet contrast, they estimate the required mean null as:

$$\langle nl \rangle (\lambda) = 1.0 \; 10^{-5} \left(\frac{\lambda}{7\,\mu\text{m}}\right)^{3.37}.$$

If such performance in terms of the mean value is achieved, its required long-term stability (at $7\,\mu$m) can be expressed as:

$$\sigma_{\langle nl \rangle}(7\,\mu\text{m}, 10\,\text{days}) \leq 3\;10^{-9}.$$

In order to obtain such a high relative stability, an instrument with only white noise is desirable, so that instability is reduced with integration time $\tau$ as $\tau^{-1/2}$.

Courtesy of our colleagues (Ollivier 1999; Alcatel 2004; Vink et al. 2003; Brachet 2005), we were able to analyse their nulling-experiment data. Unfortunately, in all cases, the power spectral density (PSD) of the null output displays a strong peak at low frequencies, i.e., a $(1/f)^\alpha$-type behaviour, with $\alpha \geq 1$. Consequently, none of these experiments show a $\tau^{-1/2}$ decrease in the standard deviation of the integrated null value, which means that the required performance cannot be obtained during very long integrations.

## 2. Principle of Optical Path Difference Dithering

OPD is one of the first quantities that needs to be controlled in an interferometric setup. Several strategies may, and in fact, should be applied simultaneously. One approach uses a separate metrological servo system, based on a laser beam following the path of the beam whose behaviour is actually under study ("scientific beam") as closely as possible. The strong points of such systems include high accuracy, monitoring as well as servo functions, high-frequency servo control, etc. The disadvantages have to do with the fact that metrology always uses monochromatic lasers whereas the scientific beam ultimately has to be a polychromatic one. Moreover, the metrological laser is very often at a wavelength outside of the working band of the experiment, which means that the OPD's and the flaws seen by the two systems may be different.

An alternative approach is dither stabilisation, implementing a servo mechanism based on the scientific beam itself. It is a standard technique in control engineering defined as "the modification of the low-frequency properties of an unstable nonlinear system by the application of a high-frequency signal in order to stabilise the system" (Gelb & Vander Velde 1968).

A classical form of this technique has been investigated by Ollivier et al. (2001), although this first experiment was inconclusive. More recently, Schmidtlin et al. (2005) have demonstrated the potential of dithering in a sequential way, obtaining good levels of stabilisation with nulling ratio around $8\,10^{-7}$ with a laser diode at 638 nm. It was this work that provided our team with decisive inspiration.

The present algorithm, however, departs in two ways from the classical system described by the cited definition (the very formulation of which is reminiscent of analogous electronic signal processing). First, the goal of the classical dithering method is to obtain a deconvoluted signal: the dithered parameter is changing continuously, and the measured signal is a convolution of the variations due to the dithering as well as to the studied system.

In our case, the experimental setup displays good stability on intermediate time scales ($t < 100$ s), and we only have to fight against long-term drifts. Our approach was, therefore, to reach and maintain a delay-line position $x_0$ corresponding to the best null simply by moving the delay line every few seconds, rapidly measuring the flux at positions $x_0 \pm \varepsilon$, and moving the delay line to a new position based on the information obtained during these excursions. Unlike Schmidtlin et al. (2005) who record data throughout the dithering cycle (i.e., their recorded data contain points measured during delay-line movement and of measurements at $x_0 \pm \varepsilon$), we chose to investigate a slightly different approach and to record data only during the final stage of each dithering cycle, corresponding to the best null.

Thus, the second difference between our procedure and the classical definition of dithering is that we do not have to perform any post-acqisition signal deconvolution. Our version of dithering simply helps to maximise the amount of time the system stays in its optimal setting; i.e. employing the dithering only when needed, with relatively long periods of unperturbed data acquisition.

The nulling ratio in the vicinity of the dark fringe's centre, with other parameters constant, can be described as a function of the phase shift $\Delta\varphi$

$$nl = \frac{I_{\min}}{I_{\max}} = \frac{1 + \cos(\pi + \Delta\varphi)}{1 + \cos(\Delta\varphi)}$$

$$nl \approx \frac{1 - 1 + \frac{\Delta\varphi^2}{2}}{1 + 1 - \frac{\Delta\varphi^2}{2}} \approx \frac{\Delta\varphi^2}{4} \quad \text{for} \quad \Delta\varphi \ll 1.$$

Bracewell's method uses achromatic phase shifters (APS) to produce a $\pi$ phase shift independent of wavelength (within a given band). The phase shift can be translated into the OPD between the two arms of the interferometer

$$\Delta\varphi(\lambda) = 2\pi\frac{x}{\lambda}$$

where $x$ is the OPD. The signal during an OPD scan can be described as tracing a parabola around the point where the nulling ratio reaches its minimum versus $\lambda$ ($x \ll \lambda$)

$$I = ax^2 + bx + c.$$

The vertex of the parabola corresponds to the deepest null obtainable at a given moment adjusting the delay line, i.e. the centre of the dark fringe. The drift in nulling performance due to OPD instability can therefore be represented as a shift of the parabola and of its vertex.

With the empirical knowledge of three points of a parabola we can calculate the position of its vertex unambiguously. Since measurement of three points of the parabola is required, the OPD has to be modified deliberately. In our experimental setup the speed of operations is limited by the speed of the delay line actuators and by the integration time required to get a sufficient S/N. This allows for compensation of slow drifts (observable on time scales of a few tens or hundreds of seconds).

If data are acquired at three different OPD's, e.g. at $x_0$, $x_0 + \varepsilon$, and $x_0 - \varepsilon$, then the position of the dark fringe, i.e. of the vertex $x_v$ of the parabola, can be calculated as

$$x_v = x_0 - \frac{I_+ - I_-}{2(I_+ + I_-) - 4I_0}\varepsilon$$

where $I_0$, $I_+$, and $I_-$ are the signal values measured at $x_0$, $x_0 + \varepsilon$, and $x_0 - \varepsilon$, respectively. This formula represents a simple recipe that can be directly implemented (Fig. 1).

## 3. Results

We have tested this method on the SYNAPSE test bench (Brachet 2005), and found it very convenient to use. It is highly efficient in finding the physical dark fringe (within one or two iterations), and reliable, driving the OPD back to its optimal value in spite of artificial perturbations.

The work of Brachet (2005) has documented two features of the SYNAPSE nulling performance. The first is the maximum rejection factor achieved. The best polychromatic (K band) performance for $\langle nl \rangle$ was $\langle nl \rangle = 2\,10^{-4}$.

The second is the nulling stability. A typical acquisition of about 200 s had a mean rejection factor of $\langle nl \rangle = 2.7\,10^{-4}$ and a standard deviation of $\sigma_{\langle nl \rangle}(200\,\text{s}) = 6\,10^{-5}$. A drift in nulling performance can be observed after, typically, 100 s, with the rejection factor gradually deteriorating to $nl \approx 10^{-3}$ after 1000 s.

Using the same experimental setup, we performed several long OPD-stabilised acquisitions. Figure 2 presents one of them,

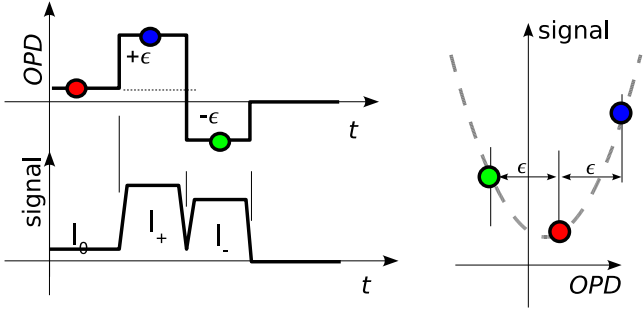

**Figure 1.** OPD dithering cycle. The top curve shows schematically the position of the delay line whereas the bottom curve is an idealised representation of the corresponding detected signal. From an initial position the dithering alters the OPD value by $+\varepsilon$(OPD) with respect to the initial position. A measurement of the flux is performed. During the next stage, the dithering algorithm again alters the OPD, this time by $-\varepsilon$(OPD) with respect to the initial position. Another measurement is performed. We thus know three points, defining a parabola (right), which allows us to calculate and reach a new working position of the OPD (the vertex of the parabola).

of duration 6 hours, with $\tau = 6.25$ s per modulation cycle, modulation amplitude $\varepsilon = 5$ nm, i.e., $\Delta\varphi = 1.5\,10^{-2}$. The fluxes in the two arms of the interferometer were balanced before the acquisition (using an adjustable semi-planar knife-edge) with an accuracy better than 0.5 percent. The detector dark current produced a lock-in amplifier signal of $(0.76 \pm 0.60)\,\mu$V, and the signal at the light fringe was $I_{max} = (8.4 \pm 1.0)$ mV.

Figure 3 shows the power-spectrum density of the nulling function, PSD($nl$). Note that no $1/f$ component can be distinguished. A comparison with data, courtesy of three other groups, in Fig. 4 makes this absence even more conspicuous. *It must be emphasised that these three experiments had the goal of achieving a low null value but not maintaining its stability.* Figure 3 also shows the standard deviation of the mean value of the nulling function over the time interval $\tau$, $\sigma_{\langle nl \rangle}(\tau)$. Note that up to $\tau \approx 600$ s, the experimental curve is consistent with the $\tau^{-1/2}$ behaviour in agreement with the flat PSD of Fig. 3 for $\nu \geq 2\,10^{-3}$ Hz. The deviation from the $\tau^{-1/2}$ behaviour that can be seen for $\tau > 600$ s in Fig. 3, is related to the peak in the PSD at $1.7\,10^{-3}$ Hz. Its origin has not been clearly established so far. It is also worth noting that $\sigma_{\langle nl \rangle}(600\,\text{s}) \approx 1.5\,10^{-5}$. Figure 4 again provides a comparison with the data, courtesy of our colleagues. The efficiency of OPD dithering in stabilising the setup stands out. This corresponds to an improvement of nulling stability with integration time. *It must be stressed, however, that we start from a very modest value of $\langle nl \rangle \approx 4\,10^{-4}$ and $\sigma_{\langle nl \rangle} \approx 2\,10^{-4}$ at $\tau \approx 3$ s, and reach $\sigma_{\langle nl \rangle} \approx 6\,10^{-6}$ for $\tau \approx 2$ h.* This last value is significantly lower than results obtained by our colleagues, e.g., the Astrium group (Flatscher et al. 2003) reached $\sigma_{\langle nl \rangle} \approx 2\,10^{-7}$ for $\tau \approx 100$ s.

## 4. Discussion and conclusion

The choices of dithering-cycle parameters are a result of several trade-offs. A limitation is imposed by the delay line's movement. The piezoelectric actuator reaches a position close to the desired one very rapidly, but then its controller takes some time, proportionate to the movement's amplitude $\varepsilon$, to stabilise the system at the new position: $\tau_{DL} \approx (90 + 6\frac{\varepsilon}{\text{nm}})$ ms. For $\varepsilon$ of 5 nm, each dithering cycle therefore implies $3\,\tau_{DL} \approx 360$ ms of waiting time.

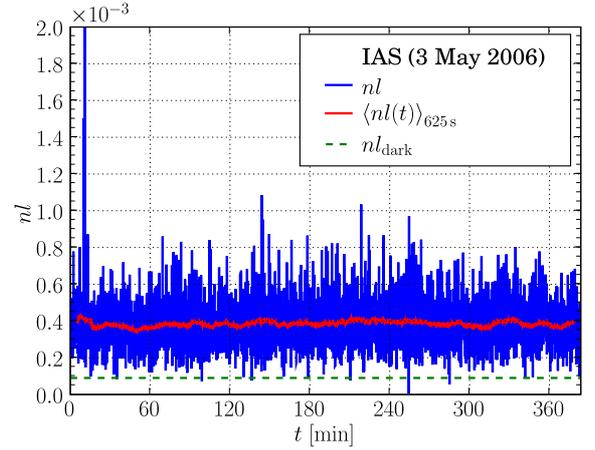

(a) Nulling ratio $nl$ from a 6 hour-long OPD dither-stabilised data acquisition

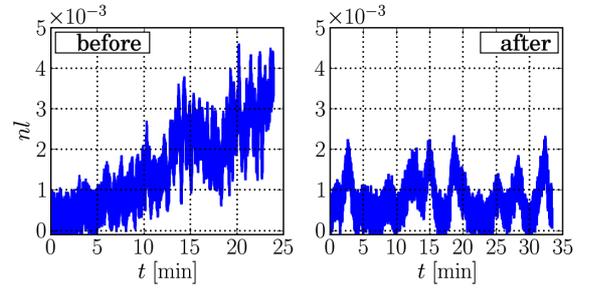

(b) Data acquisition with no stabilisation taken before and after the long stabilised run

**Figure 2.** (a) Nulling ratio (or stellar leakage) $nl(t)$ as a function of time (blue plot), with the running average calculated over 625 seconds $\langle nl(t) \rangle_{625\,\text{s}}$ overplotted (red). The nulling ratio corresponding to detector dark current (which has been subtracted from the signal in order to obtain the $nl$) was $9\,10^{-5}$. (b) Data acquisitions with no stabilisation taken immediately before and after the long stabilised run.

As for the dithering amplitude $\varepsilon$ itself, we accepted a limitation imposed by the gauge of the data acquisition system. We used a lock-in amplifier and the most straightforward solution is to maintain the same gauge for all measurements: at $x_0$, $x_0 \pm \varepsilon$ and $x_v$, alike. The flux $I_\pm(\varepsilon)$ measured at $x_0 \pm \varepsilon$ can be approximated as $I_{max}\,\Delta\varphi^2/4 = I_{max}\,(\pi\varepsilon/\lambda)^2$, whereas the flux at $x_v$, expressed as $I_{max}\,nl$, is due to a residual flux mismatch between the two arms of the interferometer, polarisation issues, and other factors contributing to nulling degradation. On the whole, we found that $\varepsilon = 5$ nm, i.e. $\Delta\varphi = 1.5\,10^{-2}$, was well suited for our purposes.

The dithering cycle with a duration of $\tau_{cyc}$ can be regarded as comprising integration time $\tau_v$ at $x_v$ and time $\tau_{cyc} - \tau_v$ when dithering actions are performed. The noise (assuming it is white noise in the relevant frequency band), measured by the standard deviation $\sigma$, decreases with the square root of the integration time. The scaling factor for $\sigma$ is therefore $\sqrt{\frac{\tau_{cyc}}{\tau_v}}$. If $\tau_v/\tau_{cyc} = 2/3$, the noise will increase by 22 %. If $\tau_v/\tau_{cyc} = 9/10$, the noise will increase by 5 %. Therefore, in our experiment we did not consider it a strong priority to increase $\tau_v/\tau_{cyc}$ beyond $\approx 2/3$.

There are two possible upper limits for the integration time $\tau_\pm$ of the flux $I_\pm$ measurements at $x_v \pm \varepsilon$. One is given by the fact that there is no reason for the corresponding SNR to be better than that of the flux $I_v$ measurements at $x_v$. This can be expressed as $\tau_\pm \leq \tau_v I_v / I_\pm = \tau_v\, 4nl/\Delta\varphi^2$ (assuming white noise only). The second upper limit for $\tau_\pm$ to be considered is given by the time $\tau_{DL}$ taken up by delay-line movements. As a rule of thumb, since it takes $\tau_{DL}$ to reach a position, there is little practical gain in reducing $\tau_\pm$ to values less than $\tau_{DL}$, which leads to $\tau_\pm \leq \tau_{DL}$. In practice, it is this latter upper limit that will be more applicable.

Although these results have to be regarded as preliminary, they nonetheless demonstrate that OPD dithering is a promising technique. The *Darwin*/TPF-I requirements (Chazelas et al. 2006) integration time of 10 min: $\langle nl \rangle = 10^{-5}$ and $\sigma_{\langle nl \rangle}(10\,\text{days}) = 3\,10^{-9}$ after a 10-day integration. Presently, we have obtained $\langle nl \rangle = 4\,10^{-4}$ and $\sigma_{\langle nl \rangle}(3\,\text{s}) = 2\,10^{-4}$ at integration times of 3 s, which improves to $\sigma_{\langle nl \rangle}(2\,\text{hrs}) = 8\,10^{-6}$ after 2 hours of integration. Since $\sigma$ in Fig. 3 is consistent with $\tau^{-1/2}$ behaviour up to $\tau \approx 500\,\text{s}$, with $\sigma_{\langle nl \rangle}(500\,\text{s}) \approx 1.5\,10^{-5}$, we extrapolate that if drifts were kept at bay we would obtain $\sigma_{\langle nl \rangle}(10\,\text{days}) = 3\,10^{-7}$ after an integration of 10 days, still two orders of magnitude short of the goal.

Progressing to the *Darwin*/TPF-I performance levels necessarily requires improving the mean null from $\langle nl \rangle = 4\,10^{-4}$ to $10^{-5}$, which has already been achieved by two of the reported experiments of our colleagues (Flatscher et al. 2003; Alcatel 2004). It also requires a better short-term stability, $\sigma_{\langle nl \rangle}(1\,\text{s}) = 8\,10^{-5}$ at an integration time of 1 s, and most importantly, a full control of drifts up to very long integration times ($\approx 10\,\text{days}$), so that $\sigma_{\langle nl \rangle}$ decreases as $\tau^{-1/2}$. Whether we can achieve such drift control with dithering when the short-term performance is improved by a factor of 40 is a crucial question that remains open.

In the near future, we will perform studies of the Synapse testbench using monochromatic laser light to improve the setup (e.g., polarisation filtering, larger flux and/or dynamic range will be required). We will also test other stabilisation techniques: metrological servo systems, and flux-balance stabilisation again with dithering (it may be interesting to compare the efficiency of this intrinsically-chromatic method with the approach of Peters et al. (2006) who have studied a deformable-mirror based compensator of both phase (OPD), and flux balance) at different wavelengths individually. The questions to be addressed are: How far can we go using dithering? How many photons do we need to sacrifice to stabilisation? In our experiment it was about a third ($\tau_v/\tau_{cyc} \approx 2/3$).

One lesson for *Darwin*/TPF-I stands out: since all means available will have to be considered if the mission's stringent performance and stability requirements are to be met and maintained in an automated space setup, the applicability, efficiency, and limitations of techniques such as OPD dithering has to be studied. In addition, these methods have to be considered as an inherent part of the system's design, with a possible impact upon conceptual choices.

*Acknowledgements.* This work was supported by *Centre National d'Etudes Spatiales* (*action R&T,* n. R-506/SU-002-022) and *Agence Nationale de la Recherche* (*thème blanc,* n. 5A 0617).

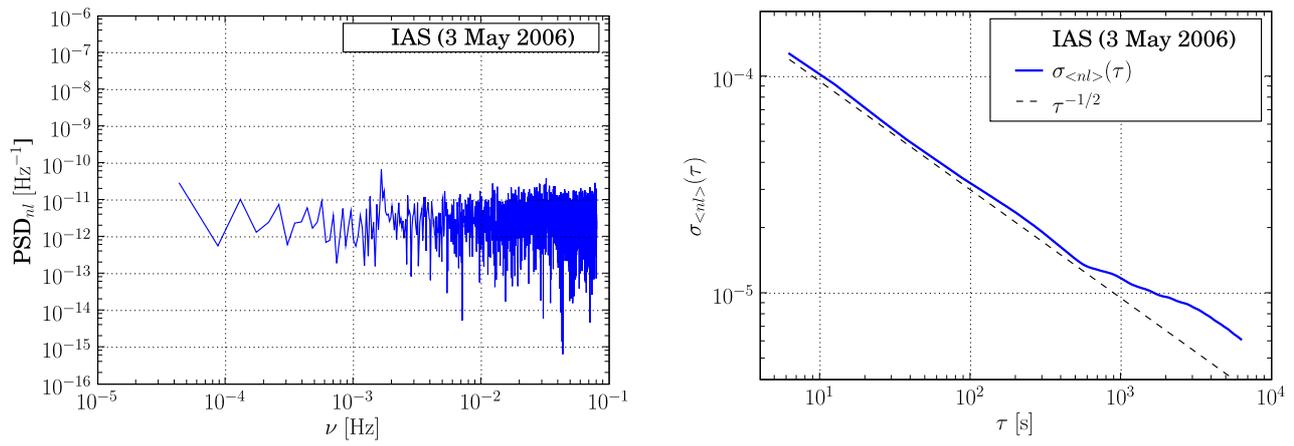

**Figure 3.** Left: Power spectrum density of the nulling function, PSD($nl$). Note that $1/f$ component is negligible. Right: Standard deviations of the running average of the nulling function over the time interval $\tau$, $\sigma_{\langle nl \rangle}(\tau)$ (curve above). A (displaced) $\tau^{-1/2}$ function is shown for comparison (line below). Note that up to $\tau \approx 500$ s, the experimental curve is consistent with the $\tau^{-1/2}$ behaviour, with $\sigma_{\langle nl \rangle}(500\,\text{s}) \approx 1.5\,10^{-5}$.

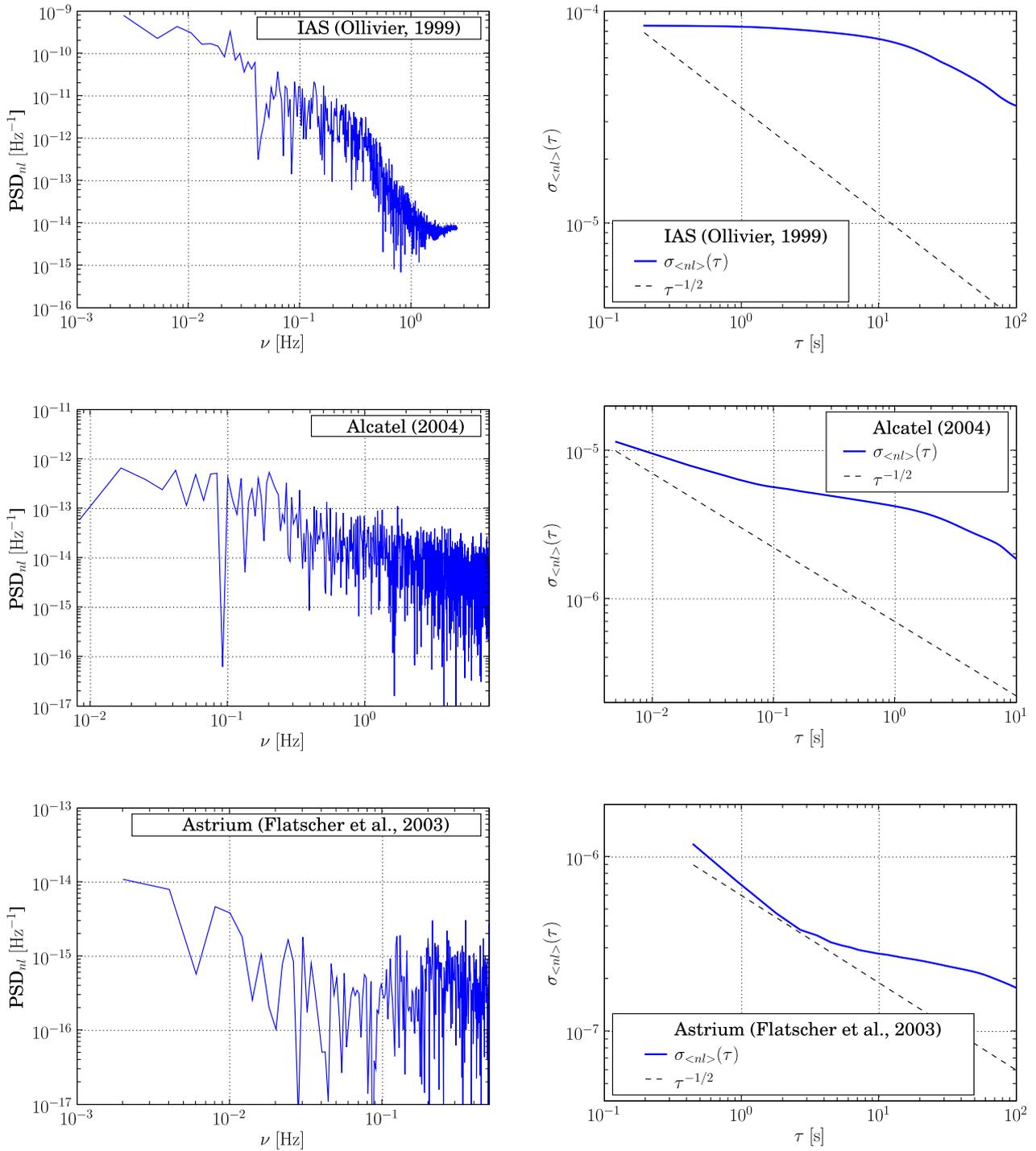

**Figure 4.** Comparison with three other experiments. Left column: Power spectrum distributions. Right column: Standard deviation of the running average of $nl$ over time $\tau$, $\sigma_{\langle nl\rangle}(\tau)$. Top: Results from the nulling experiment Ollivier (1999); Ollivier et al. (2001). The increase in the PSD for low values of $\nu$ is clear. In the range of 0.3-0.01 Hz the PSD is approximately $\nu^{-1.35}$, a behaviour close to the "classical" $1/f \equiv 1/\nu$ behaviour. In the frequency range investigated by that experiment, $\sigma$ decreases with $\tau$ but more slowly than $\tau^{-1/2}$, which is typical of PSDs with $1/f$-like components. Centre: Results from the nulling experiment by Alcatel (2004), using a laser diode at $\sim 1.57\mu m$ and an integrated optics recombiner. The increase of the PSD at low frequencies is clear. From 0.1 to 1 Hz the curve goes approximately as $1/\nu$ (courtesy Alcatel Space Industry). Bottom: Results obtained by Flatscher et al. (2003). The duration of the experiment is the longest of the three. The null curve between 1000 and 1500 s is selected (Chazelas et al. 2006) to compute the PSD because it has the best qualities. A low-frequency increase in the PSD is present. From 0.01 to 0.1 Hz it is approximately $1/\nu$ (courtesy Astrium, Germany).